\documentclass[]{aa}
\usepackage{epsfig} 
\def\nerr#1#2{{\,+{#1}}\,-{#2}}
\begin{document}
\thesaurus{3 (11.02.2, 13.07.2)}
\title{Observations of Mkn 421 during 1997 and 1998 
in the energy range above 500 GeV
with the HEGRA stereoscopic Cherenkov telescope 
system}
\author{F.A. Aharonian\inst{1},
A.G. Akhperjanian\inst{7},
M. Andronache\inst{4},
J.A.~Barrio\inst{2,3},
K.~Bernl\"ohr\inst{1,}$^*$,
H. Bojahr\inst{6},
I. Calle\inst{3},
J.L. Contreras\inst{3},
J. Cortina\inst{3},
A. Daum\inst{1},
T. Deckers\inst{5},
S. Denninghoff\inst{2},
V. Fonseca\inst{3},
J.C. Gonzalez\inst{3},
N. G\"otting\inst{4},
G. Heinzelmann\inst{4},
M. Hemberger\inst{1},
G. Hermann\inst{1,}$^\dag$,
M. He{\ss}\inst{1},
A. Heusler\inst{1},
W. Hofmann\inst{1},
H. Hohl\inst{6},
D. Horns\inst{4},
A. Ibarra\inst{3},
R. Kankanyan\inst{1,7},
J. Kettler\inst{1},
C. K\"ohler\inst{1},
A. Konopelko\inst{1,}$^\S$,
H. Kornmeyer\inst{2},
M. Kestel\inst{2},
D. Kranich\inst{2},
H. Krawczynski\inst{1},
H. Lampeitl\inst{1},
A. Lindner\inst{4},
E. Lorenz\inst{2},
N. Magnussen\inst{6},
O. Mang\inst{5},
H. Meyer\inst{6},
R. Mirzoyan\inst{2},
A. Moralejo\inst{3},
L. Padilla\inst{3},
M. Panter\inst{1},
D. Petry\inst{2,}$^\ddag$,
R. Plaga\inst{2},
A. Plyasheshnikov\inst{1,}$^\S$,
J. Prahl\inst{4},
G. P\"uhlhofer\inst{1},
G. Rauterberg\inst{5},
C. Renault\inst{1,}$^\#$,
W. Rhode\inst{6},
A. R\"ohring\inst{4},
V. Sahakian\inst{7},
M. Samorski\inst{5},
M. Schilling\inst{5},
F. Schr\"oder\inst{6},
W. Stamm\inst{5},
H.J. V\"olk\inst{1},
B. Wiebel-Sooth\inst{6},
C. Wiedner\inst{1},
M. Willmer\inst{5},
W. Wittek\inst{2}}
\institute{Max Planck Institut f\"ur Kernphysik,
Postfach 103980, D-69029 Heidelberg, Germany \and
Max Planck Institut f\"ur Physik, F\"ohringer Ring
6, D-80805 M\"unchen, Germany \and
Universidad Complutense, Facultad de Ciencias
F\'{i}sicas, Ciudad Universitaria, E-28040 Madrid, Spain 
\and
Universit\"at Hamburg, II. Institut f\"ur
Experimentalphysik, Luruper Chaussee 149,
D-22761 Hamburg, Germany \and
Universit\"at Kiel, Institut f\"ur Experimentelle und Angewandte Physik,
Leibnizstra{\ss}e 15-19, D-24118 Kiel, Germany\and
Universit\"at Wuppertal, Fachbereich Physik,
Gau{\ss}str.20, D-42097 Wuppertal, Germany \and
Yerevan Physics Institute, Alikhanian Br. 2, 375036 Yerevan, 
Armenia\\
\hspace*{-4.04mm} $^*\,$ Now at Forschungszentrum Karlsruhe,
P.O. Box 3640, D-76021 Karlsruhe, Germany\\
\hspace*{-4.04mm} $^\dag\,$ Now at Enrico Fermi Institute,
The University of Chicago, 933 East 56th Street,
Chicago, IL 60637, U.S.A.\\
\hspace*{-4.04mm} $^\ddag\,$ Now at Universidad Aut\'{o}noma de 
Barcelona,
Instituto de F \'{\i}sica d'Altes Energies, E-08193 Bellaterra, Spain\\
\hspace*{-4.04mm} $^\S\,$ On leave from  
Altai State University, Dimitrov Street 66, 656099 Barnaul, Russia\\
\hspace*{-4.04mm} $^\#\,$ Now at LPNHE, Universit\'es Paris VI-VII, 4
place Jussieu, F-75252 Paris Cedex 05, France
}
\mail{Henric Krawczynski, \\Tel.: (Germany) +6221 516 471,\\
email address: Henric.Krawczynski@mpi-hd.mpg.de}
\offprints{Henric Krawczynski}
\date{Received ---; accepted ---}
\authorrunning{F. Aharonian et al.}
\titlerunning {TeV characteristics of Mkn~421}
\maketitle
\begin{abstract}
Since its commissioning in fall 1996,
the stereoscopic system of Imaging Atmospheric 
Cherenkov Telescope (IACTs) of HEGRA with an
energy threshold of 500 GeV, an angular resolution of 
0.1$^\circ$ and an energy resolution of 20\% per 
individual photon, and an energy flux 
sensitivity $\nu\,$F$_\nu$ at 1 TeV of 10$^{-11}$ 
$\rm erg\, cm^{-2}\, s^{-1}$ (S/N = 5$\sigma$) for
one hour of observation time has been 
used to monitor the BL Lac object Mkn 421
on a regular basis.
In this letter, we report detailed temporal and spectral
information about the TeV characteristics of 
Mkn 421 in 1997 and 1998.
We study the light curve, the shortest time scales
of flux variability, the differential spectra on diurnal basis
for several days with good $\gamma$-ray statistics and
the time averaged energy spectrum. Special emphasis will be put
on presenting the data taken during the world-wide April 1997 
multiwavelength campaign.
We compare the Mkn 421 results with the results obtained for 
the BL Lac object Mkn 501 and discuss possible implications for
the emission mechanism and the Diffuse Intergalactic
Background Radiation.
\end{abstract}
\sloppy
\section{Introduction}
The advent of the {\it Compton Gamma Ray Observatory} and the EGRET 
instrument on board has led to the spectacular discovery of intense 
high energy $\gamma$-ray radiation from as many as $\sim$65 
Active Galactic Nuclei (AGN) in the 0.1 to 10 GeV energy range 
(Hartman et al. \cite{Hart:99}). The recorded sources
belong mainly to the blazar class, i.e.\ BL Lac objects,
Optically Violent Variables (OVVs), and highly polarized
quasars, all radio-loud sources which show variability on short
time scales in most of the observed frequency bands.
The power emitted in $\gamma$-radiation frequently
dominates the power radiated by the source 
(see e.g.\ the spectral energy distributions shown by
Fossati et al.\ \cite{Foss:98}).
It is widely believed, that the non-thermal $\gamma$-ray emission
is produced in a relativistic jet
by a population of  electrons, emitting
synchrotron radiation at longer wavelengths and $\gamma$-rays
in Inverse Compton (IC) processes of the highest energy electrons
with lower energy seed photons 
(see for recent reviews Coppi \cite{Copp:97}; Sikora \cite{Siko:97};
Ulrich et al.\ \cite{umu:97}).
The origin of the IC seed photons has not yet been established, debated 
possibilities include a target photon population dominated 
by lower energy synchrotron photons 
(the Synchrotron Self Compton mechanism, see e.g.\ 
Bloom \& Marscher \cite{Bloo:93};
Ghisellini et al.\ \cite{GhisMD:1996}; 
Mastichiadis \& Kirk \cite{Mast:97}), or dominated by external
photons, e.g.\ by radiation from the nuclear continuum scattered or reprocessed 
in the broad-line regions
 (Sikora, Begelmann \& Rees \cite{Siko:94};
Blandford \& Levinson \cite{Blan:95}), or
by  accretion disc photons (Dermer \& Schlickeiser \cite{Derm:94}).
Although these possibilities have been studied in great detail,
models where hadrons are the particles primarily accelerated 
(see e.g.\ Mannheim et al.\ \cite{Mann:93}) can not be ruled out yet.
Clearly, the study of the emission process could yield
crucial clues regarding the structure of the jets leading ultimately 
to an understanding of the mechanisms of energy extraction operating
in the surrounding of the central supermassive object.

In this paper, we report on the 1997 and 1998 observations of 
the BL Lac object Mkn 421 (redshift $z$= 0.031) 
with the IACT system of HEGRA (Daum et al.\ \cite{Daum:97}).
The system provides an energy threshold of 500 GeV, an angular 
resolution of  0.1$^\circ$ and an energy resolution of 20\% for
individual photons, and an energy flux 
sensitivity $\nu\,$F$_\nu$ at 1 TeV of 10$^{-11}$ 
$\rm erg\, cm^{-2} s^{-1}$ (S/N = 5$\sigma$) for one hour of 
observation time. It has been used to monitor Mkn 421 on a regular basis 
ever since its commissioning in fall 1996 to gain
detailed temporal and spectral information.

TeV observations yield information about the emitting particles
with very high energies, complementary to the information from
longer wavelengths.
Due to the extreme conditions needed for the
production of TeV photons, TeV blazars could be extremely revealing
laboratories for the understanding of these sources in general.
It should be pointed out,  that the TeV radiation is expected to 
be partially absorbed by the Diffuse Extragalactic Background Radiation 
(DEBRA)
(e.g.\ Gould \& Schr\'{e}der \cite{Goul:65}; Stecker, de Jager \&
Salomon \cite{Stec:92}). While on the one hand the DEBRA extinction 
opens unique possibilities to infer information about the DEBRA 
density in the largely unconstrained 0.5 to  100~$\mu$m 
wavelength region (see Aharonian et al. \cite{Ahar:99b}, called 
{\it Paper 2} in the following), 
on the other hand it renders the understanding of the source 
more difficult (e.g. Coppi \& Aharonian \cite{Copp:99};
Bednarek \& Protheroe \cite{Bedn:99}).

The BL Lac Mkn 421 was the first extragalactic object to be 
discovered as a TeV emitter
(Punch et al.\ \cite{Punc:92}; Petry et al.\ \cite{Petr:96}).
The early HEGRA observations during the 1994/1995 observation period
showed, that the spectrum of Mkn 421 
was significantly softer than the Crab spectrum
(Petry  et al.\ \cite{Petr:96}).
While the mean TeV flux level of Mkn 421 is about 0.5 Crab units 
(Petry  et al.\ \cite{Petr:96}), flares exceeding five Crab units
on timescales as short as 15 minutes have been observed
(Gaidos et al.\ \cite{Gaid:96}).

The second BL Lac object to be discovered as TeV source 
(Quinn et al.\ \cite{Quin:96}; Bradbury et al.\ \cite{Brad:97}),
Mkn 501 (redshift $z$= 0.034),  
underwent a major outburst in X-ray and in the TeV energy range during 
1997 (see e.g.\ Protheroe et al.\ \cite{Prot:98}).
Detailed accounts of the results obtained with the HEGRA IACTs
are given in (Aharonian et al.\ \cite{Ahar:99a}, called 
{\it Paper 1}
in the following), Paper 2, and in (Aharonian et al.\ \cite{Ahar:99c}).
The source emission showed strong variability with differential 
fluxes at 1~TeV from a fraction to 10 times the Crab flux. The HEGRA IACT system was 
used to determine diurnal differential spectra for more than 60 
individual days.
Surprisingly no correlation between absolute flux and spectral 
shape was
found. The differential time averaged photon spectrum was well 
described by 
a power law  with an exponential cutoff $d$N/$d$E $\propto\,\rm (E/1 
TeV)^{-1.9}\,
\exp{(-E/6.2 TeV)}$ up to energies of $\sim$ 20 TeV.
Due to the similarity in redshifts of Mkn 421 and Mkn 501 ($z$ = 0.031 
and 0.034
respectively) the intergalactic extinction of the TeV photons caused 
by the 
DEBRA should be very similar for both objects and a comparison of both 
spectra is of 
utmost interest.

The paper is organized as follows. 
In Sect.\ \ref{HEGRA} the IACT system of HEGRA is introduced, and
in Sect.\ \ref{DC} the data sample 
and the analysis techniques are described.
The results concerning the time averaged and the diurnal 
properties, as well as
the search for variability within individual nights
are presented in Sect.\ \ref{RE}.
In Sect.\ \ref{SU} the results are summarized and discussed.
\section{The HEGRA Cherenkov telescope system}
\label{HEGRA}
The HEGRA collaboration operates six imaging atmospheric Cherenkov telescopes
located on the Roque de los Muchachos on the Canary island of La Palma, 
at 2200~m above sea level. 
A prototype telescope (CT1) started operation in 1992 and has undergone 
significant hardware upgrades since then (Mirzoyan et al.\ \cite{Mirz:94};
Rauterberg et al.\ \cite{Raut:95}). 
This telescope continues to operate as an independent instrument and the 
Mkn 421 data taken with this telescope will be presented elsewhere.
The stereoscopic system of Cherenkov telescopes 
consists of five telescopes (CT2 - CT6), and has been taking data since 1996, 
initially with three and four telescopes, and since fall 1998 as a complete 
five-telescope system. Four of the telescopes (CT2, CT4, CT5, CT6) 
are arranged in the corners of a square with roughly 100 m side length, 
and one telescope (CT3)
is located in the center of the square. During 1997 and 1998, when the data 
discussed in this paper were taken, CT2 was still used as stand alone detector.

The telescopes have an 8.5 m$^2$ tessellated reflector, focusing the 
Cherenkov light onto a camera with 271 Photomultipliers (PMTs), covering 
a field of view of $4.3^\circ$ in diameter. A telescope is triggered when the 
signal in at least two adjacent PMTs exceeds an amplitude of 10 
(before June 1997)
or 8 (after  June 1997)
photoelectrons; in order to trigger the IACT system 
and to initiate the readout 
of data, at least two telescopes have to trigger 
simultaneously. Typical trigger 
rates are in the 10-16 Hz range. The PMT signals are digitized and recorded 
by 120 MHz Flash-ADCs. The telescope hardware is
described by Hermann (\cite{her:95}); 
the trigger system and its performance are 
reviewed by Bulian et al.\ (\cite{bul:98}); the calibration of the
detector sensitivity is described in (Fra{\ss} et al.\ \cite{Fras:97}),
and the determination of the correct telescope pointing with
35 arc sec accuracy is described in (P\"uhlhofer et al.\ \cite{pul:97}).
\section{Data sample and analysis tools}
\label{DC}
The analysis described in the following is based on 165 hours of 
data
acquired between January 1st, 1997 and May 27th, 1998 with Mkn 421 altitudes
larger than 45$^\circ$ above the horizon.
Only data taken with all four IACTs during 
excellent weather conditions, i.e.\ 
clear sky and humidity below 90\%, have been used.
Furthermore, data runs with a mean Cosmic Ray rate 
deviating by more than 21\% 
from the expected value 
or with a mean width parameter per telescope deviating by more than 
6\% from the expected value 
were excluded from the analysis.

The Monte Carlo simulations used for the data analysis 
are described in detail in (Hemberger \cite{Hemb:98}; Konopelko et al.\ 
\cite{Kono:99}).
Changes of the trigger conditions, the mirror alignment and 
reflectivities as well as
in the PM sensitivities divide the data sample in
eight phases of approximately uniform detector performance.
For each of these eight phases, the detector simulations have been
adjusted to describe as closely as possible the hardware
performance; hereby, information about the mirror alignment from
point runs, information about the mirror reflectivities and
Photomultiplier sensitivities from the measured Cosmic Ray rates, and
the measured single pixel trigger characteristics have been
taken into account (see Paper 1, 2 and references therein).
The Monte Carlo fine tuning and the data quality
have intensively been tested using data from Mkn 501,
Mkn 421, and from the Crab nebula, allowing to effectively
test all response functions relevant for the determination of
$\gamma$-ray fluxes and $\gamma$-ray energy spectra (see Paper 1 \& 2, and
Hofmann \cite{hof:97}).
The current status of the Crab analysis with the 
HEGRA telescope system is described in (Konopelko et al.\ \cite{kon:98}).

The standard analysis methods described in detail in 
Paper 1 and 2 have been used.
Thus, spectral studies have been performed with loose selection cuts, 
i.e.\ with a cut on the angular distance $\Theta$ 
of the reconstructed direction from the 
Mkn 421 direction $\Theta^2<(0.22^\circ)^2$ and a cut on 
the parameter ``mean scaled width'' with a cut value of 1.2. 
Loose cuts do not give an optimal signal to noise ratio but
guarantee small systematic errors on the energy dependent cut
efficiencies.
The analysis uses an extended OFF-region for
minimizing the statistical errors of the background estimate.
A ring segment (of 180$^\circ$ opening angle), 
from 0.3$^\circ$ to 0.7$^\circ$ distance from the camera center, 
at the opposite side of the ON-region, has been chosen.
The energy reconstruction required two IACTs within 200 m from the shower 
axis,
each with  more than 40 photoelectrons per image and a ``distance'' parameter of
smaller than 1.7$^\circ$. Additionally, a minimal stereo angle 
larger than 20$^\circ$ was required. 

The methods for reconstructing the $\gamma$-ray
energy spectra, as well as the description of the main sources of systematic
uncertainties have been discussed in detail in Paper 1 and 2.
The systematic errors can be divided into two parts, into
an error on the absolute energy scale, and into additional errors
on the shape of the spectrum (see Paper 2).
The error on the absolute energy scale amounts to 15\% and is dominated by
a 10\% uncertainty 
of the detector sensitivity. Additional uncertainties arise
mainly due to the incomplete knowledge of the atmospheric 
conditions during the observations.
The error on the shape of the spectrum is considerable in the threshold 
energy region, i.e. at energies below 1 TeV. The error is dominated by
uncertainties in the trigger behavior of the IACTs and by uncertainties
of the reconstruction of $\gamma$-ray energy spectra based on
Monte Carlo simulations at discrete zenith angles.
Integral fluxes above a certain energy threshold are 
obtained by integrating the differential energy spectra above the 
threshold energy, rather than by simply scaling detection rates.
By this means the integral fluxes can be computed without
assuming a certain source energy spectrum.

For the data up to zenith angles of 30$^\circ$,
comprising altogether 126 h of data, spectral results are presented 
for energies above 500 GeV; for the 39~h of data with 
zenith angles between 30$^\circ$ and 45$^\circ$, 
an energy threshold of 1 TeV has been chosen.
If not stated otherwise, the results presented in the following 
are derived from the data with zenith angles below 30$^\circ$.
\section{Experimental results}
\label{RE}
\subsection{Time averaged properties}
\begin{figure}
\hspace*{1cm}
\resizebox{6.2cm}{!}{\includegraphics{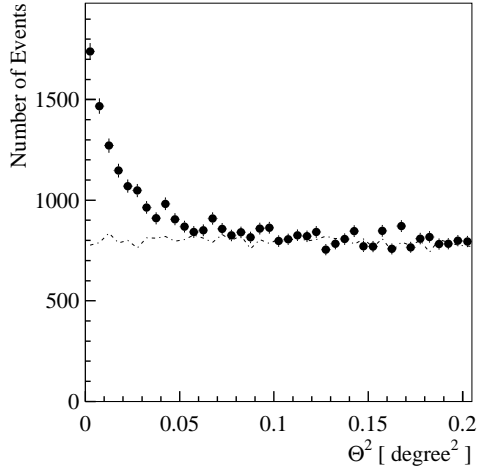}}
\caption{\label{excess} \small The Mkn 421 excess from 1997--1998. The full circles
show the 
events in the 
ON-region, the dashed line the events in the OFF-region (background region) 
after a loose ``mean scaled width'' cut and a software threshold
of at least 40 photoelectrons in two or more telescopes.}
\end{figure}

\begin{figure}
\vspace*{-0.5cm}
\resizebox{\hsize}{!}{\includegraphics{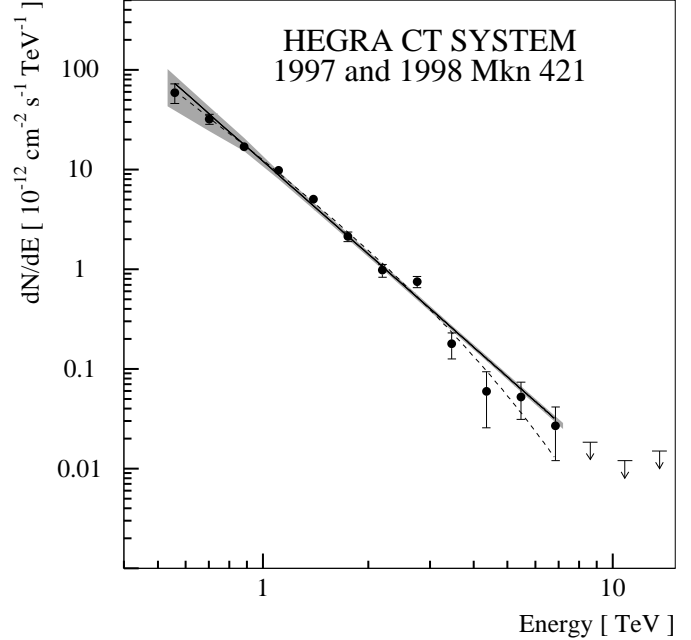}}
\vspace*{-0.5cm}
\caption{\label{spec01} \small The time averaged 1997--1998 Mkn 421 energy spectrum.
Vertical errors bars indicate statistical errors.
The hatched area gives the estimated
systematic errors, except the 15\% uncertainty on the absolute energy scale. 
The solid line shows a power law fit, 
the dashed line a power law with an exponential cut off.
Upper limits are at 2$\sigma$ confidence level.
}
\end{figure}

The total excess from Mkn 421 is shown in Fig.\ \ref{excess}. 
Using an angular cut of $\Theta^2<(0.22^\circ)^2$ the significance
of the excess (($N_{on}-N_{off})/\sqrt{N_{on}+N_{off}}\,$)
is 16$\sigma$ before image cuts and 26$\sigma$ 
after the image cut.
The total number of recorded $\gamma$-rays is about 8000 
at hardware threshold.
The time averaged energy spectrum, determined from the data with
zenith angles smaller than 30$^\circ$ is shown in
Fig.\ \ref{spec01}.
The error bars show the statistical errors, the hatched
region shows the systematic error on the shape of the spectrum described above. 
Additionally, there is a 15\% error on the absolute energy scale.
From the lowest energies of 500 GeV to the highest energies detected at
about 7 TeV, the spectrum can be described by a power law
(Fig.\ \ref{spec01}, solid line) 
\[
\frac{d\textrm{N}}{d\textrm{E}}\,\,\rm 
=\,\,(12.1 \, \pm0.5_{stat} \, \pm4.3_{syst})\, \,10^{-12}\,\,\times
\]
\begin{equation}
\hspace*{0.5cm}
\rm (E/TeV)^{(-3.09\,\pm0.07_{stat}\, \pm 0.10_{syst})}\,\,
cm^{-2} s^{-1} TeV^{-1}\,\,\,.
\label{pow} 
\end{equation}
The 36\% systematic error on the flux at 1 TeV takes into account the error
on the shape of the spectrum and the 15\% error on the energy scale. 
The $\chi^2$-value of the fit, computed with the statistical errors only, 
is 23.5 for 10 degrees of freedom, corresponding to a 
chance probability for larger $\chi^2$-values of 1\%.
Taking into account the systematic errors on the shape of the spectrum,
the result is surely compatible with a pure power law spectrum.
Note however, that a power law with an exponential cutoff: 
\begin{equation}
d\textrm{N}/d\textrm{E} \rm \,\,\propto \,\, (E/TeV)^{(-2.5 \pm 0.4_{stat})}
exp(-E/E_0) 
\label{exp}
\end{equation}
with E$_0$~=~2.8$\left({\raisebox{-0.5ex}{$\stackrel{+{2.0}}{\scriptstyle-{0.9}}$}}
\right)_{\rm stat}$~TeV,
fits the data equally well (Fig.\ \ref{spec01}, dashed line).
Here the $\chi^2$-value, also computed with the statistical errors only, 
is 16.8 for 9 degrees of freedom with a chance 
probability for larger $\chi^2$-values of 5\%.
The evaluation of the data from 30$^\circ$ to 45$^\circ$ zenith angles 
(39~h) gives similar results, although the statistics are very poor; 
the power law fit from 1 TeV to 5 TeV yields a differential spectral index of 
-3.16 $\pm0.3_{\rm stat}$.

The statistics do not permit to decide whether the energy spectrum changed
from 1997 to 1998: 
for 1997, a differential spectral index of -3.28 $\pm$0.20$_{\rm stat}$ 
is computed,
and for 1998 it is -3.00 $\pm$ 0.05$_{\rm stat}$.
\begin{figure}
\vspace*{-1.5cm}
\resizebox{\hsize}{!}{\includegraphics{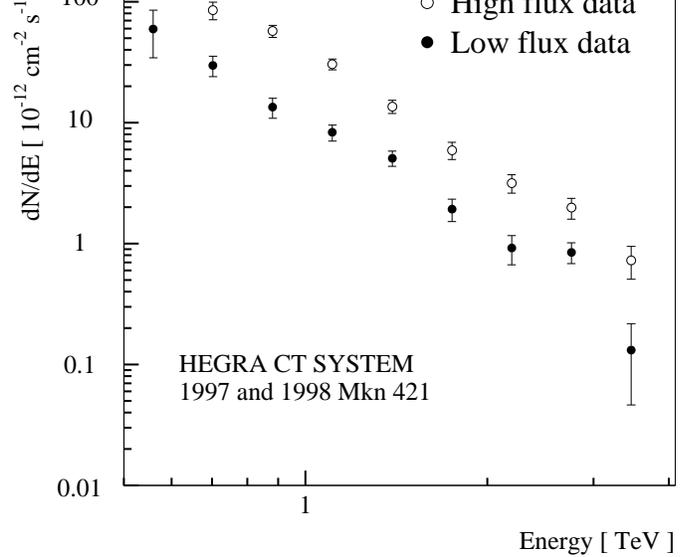}}
\caption{\label{corr1} \small The figure shows the energy spectrum for all days
with a time averaged differential flux at 1 TeV, 
between 5$\cdot\,10^{-12}$~$\rm cm^{-2} s^{-1} TeV^{-1}$
and 15$\cdot\,10^{-12}$~$\rm cm^{-2} s^{-1} TeV^{-1}$, 
and above 30$\cdot\,10^{-12}$~$\rm cm^{-2} s^{-1} TeV^{-1}$,
lower and upper curve respectively. 
(statistical errors only)}
\end{figure}

A possible correlation between absolute flux
and spectral shape has been searched for by determining 
the time averaged energy spectrum for all days with a differential 
flux at 1 TeV, 
between {\small 5$\cdot\,10^{-12}$~$\rm cm^{-2} s^{-1} TeV^{-1}$}
\clearpage
and 15$\cdot\,10^{-12}$~$\rm cm^{-2} s^{-1} TeV^{-1}$, 
and above 30$\cdot\,10^{-12}$~$\rm cm^{-2} s^{-1} TeV^{-1}$.
The first data sample contains 43~h of data, 
the second one contains  13~h of data.
The two spectra are shown in Fig.\ \ref{corr1}
for all bins with reasonable statistical errors. 
A power law fit to the high state data gives the differential spectral 
index of -2.98$\pm$ 0.09$_{\rm stat}$, and for the low state data 
we obtain -3.02$\pm$0.12$_{\rm stat}$,
not indicating any difference between the two spectra.
\subsection{Diurnal results}
\begin{figure}
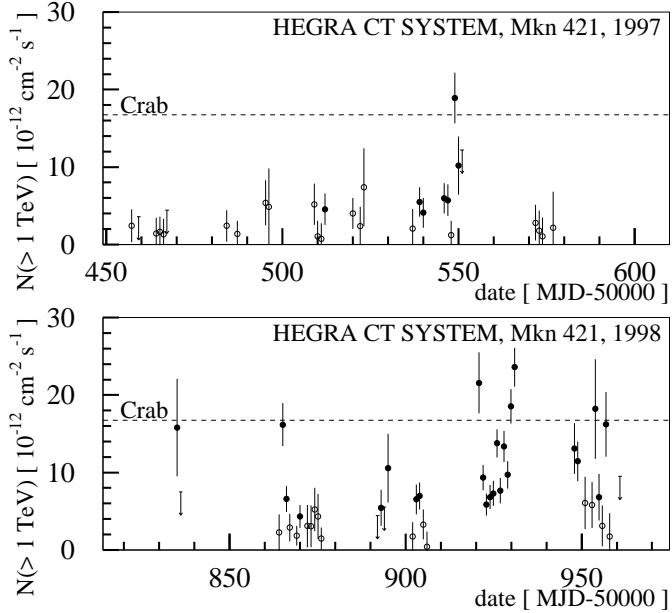

\resizebox{\hsize}{!}{\includegraphics{421.lc1.epsi}}
\resizebox{\hsize}{!}{\includegraphics{421.lc2.epsi}}
\vspace*{-0.5cm}
\caption{\label{light curve} \small The 1997 (upper panel) and 1998 (lower panel)
Mkn 421 light curve, i.e., 
the integral fluxes above 1 TeV as a function of date. 
The upper light curve starts at MJD 50449 (1.1.1997) and
ends at MJD 50610 (11.6.1997). The lower light curve starts at MJD
50814 (1.1.1998) and ends at MJD 50975 (11.6.1998). 
The full symbols show detections with at least 2$\sigma$ significance and
the open symbols are detections with less than 2$\sigma$ significance.
The upper limits are 2$\sigma$ confidence level. (statistical errors only; see text for systematic errors)}
\vspace*{+0.5cm}
\end{figure}
\begin{figure}
\resizebox{\hsize}{!}{\includegraphics{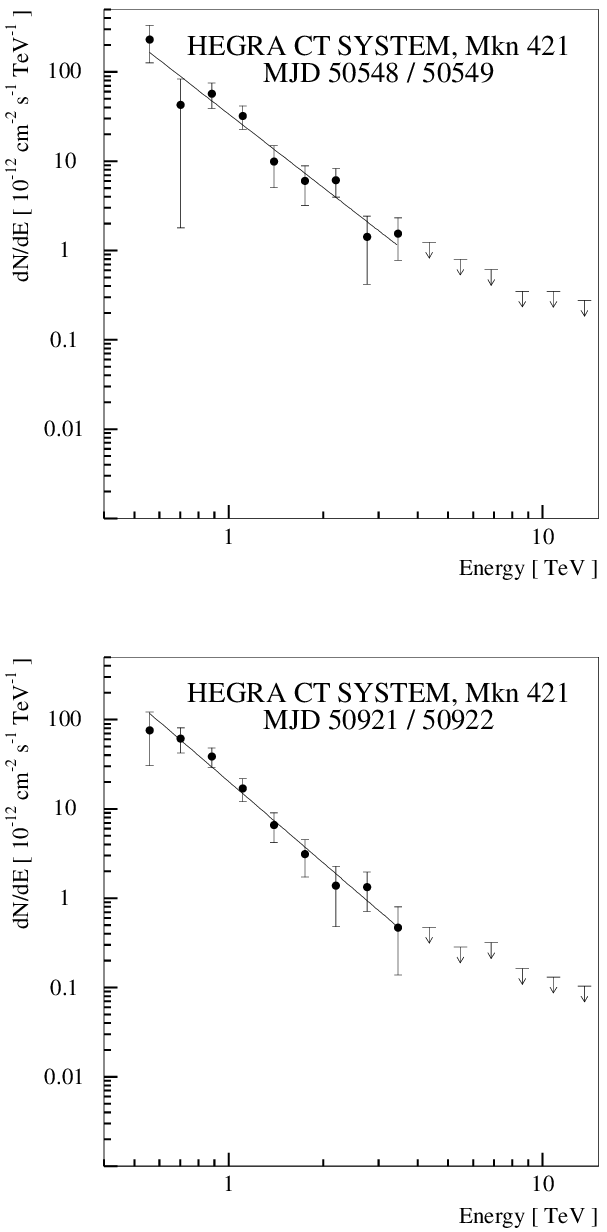}}
\caption{\label{days} \small Two diurnal Mkn 421 energy spectra.
Upper limits are at 2$\sigma$ confidence level.
(statistical errors only; systematic errors as for Fig.\ \protect\ref{spec01})}
\end{figure}

Figure \ref{light curve} shows the 1997 and 1998 Mkn 421 light curve. 
Only statistical errors are shown. The systematic
error on the integral flux, dominated by the 15\% uncertainty in the
energy scale, amounts to a 40\% relative uncertainty.
Mkn 421 was highly variable during 1997 and 1998; the fluxes varied from
an emission of a fraction of one Crab level to more than one Crab 
level.
Most of the flares seemed to last one or two days. 
In 1998, from April 17th to April 28th (MJD 50920 - MJD 50931) 
a high state lasting for approximately ten days was observed.
This flare coincided with a world-wide multiwavelength campaign lasting
approximately from April 10 to April 30 (Takahashi et al.\ \cite{Taka:99}; Urry \cite{Urry:99}).
A more detailed discussion of the HEGRA data from this campaign is given
in the next subsection.
During the whole 1997--1998 period no flares with fluxes in excess of two Crab levels
were observed.

For days with long observation times or with strong source activity,
diurnal differential spectra can be obtained. Two examples of spectra
are shown in Fig.\ \ref{days}. The first spectrum was acquired during
1.6 hours and the second one during 3.4 hours. These diurnal spectra
are statistics limited to energies below 4 TeV.
In Table \ref{strong} the results of power law fits are summarized for all
days with good statistics (3 sigma excess above 1 TeV).
No highly significant evidence for spectral variability has been found;
the diurnal differential spectral indices are statistically 
consistent with the mean index of -3.1.
The two most significant deviations from the mean index were found for
April 22nd/23rd, 1998 (MJD 50925/50926) and April 27th/28th, 1998 
(MJD 50930/50931), where the spectral indices of -2.71 and -2.82 
deviated by 2.1$\sigma$ and $1.8\sigma$ from the mean value, respectively.
\renewcommand{\arraystretch}{1.07}
\begin{table}[t]
\caption{\scriptsize HEGRA results for Mkn 421 on diurnal basis (1997 and 1998).\label{strong}
Only the statistical errors are given. The systematic error on the flux at 
1 TeV is about 40\% and the systematic error on the spectral 
indices is about 0.10.}
{\scriptsize
\begin{center}
\begin{tabular}{llrrr}
\hline
\tiny
&&&&\\
{\small start$^a$}&
{\small $\Delta t$ $^b$}&
{\scriptsize $\mathrm{d}\Phi/\mathrm{d}E\,(1\,\rm TeV)$$^c$} &
{\small spectral index}&
{\scriptsize $\Phi(>1\,{\rm TeV})$$^d$}\\
\tiny
&&&&\\
\hline
&&&&\\
50545.9181 &  2.67 &13.48\nerr{3.28}{3.06} & -3.87\nerr{0.47}{0.51} &  5.95 +-  1.96\\
50548.9648 &  1.61 &33.51\nerr{4.23}{4.34} & -2.71\nerr{0.21}{0.26} & 18.90 +-  3.28\\
50865.0528 &  1.61 &26.42\nerr{3.93}{3.87} & -2.81\nerr{0.27}{0.32} & 16.18 +-  2.78\\
50866.0482 &  2.57 & 9.07\nerr{2.43}{2.46} & -2.89\nerr{0.48}{0.55} &  6.57 +-  1.67\\
50902.9772 &  2.06 & 4.91\nerr{2.10}{2.03} & -2.05\nerr{0.48}{0.69} &  6.52 +-  1.93\\
50903.9649 &  2.59 & 9.62\nerr{2.58}{2.75} & -3.13\nerr{0.41}{0.63} &  6.97 +-  1.76\\
50920.8784 &  0.97 &32.85\nerr{5.64}{5.36} & -2.66\nerr{0.28}{0.33} & 21.58 +-  3.94\\
50921.8776 &  3.44 &20.02\nerr{2.53}{2.24} & -3.05\nerr{0.21}{0.25} &  9.34 +-  1.63\\
50922.8790 &  3.01 & 8.55\nerr{2.08}{2.20} & -2.63\nerr{0.33}{0.46} &  5.85 +-  1.41\\
50923.8785 &  2.94 &12.70\nerr{2.48}{2.69} & -3.22\nerr{0.43}{0.44} &  6.81 +-  1.55\\
50924.8804 &  3.21 & 9.43\nerr{2.30}{2.28} & -2.88\nerr{0.36}{0.51} &  7.28 +-  1.61\\
50925.8812 &  3.12 &22.11\nerr{2.30}{2.48} & -2.71\nerr{0.15}{0.18} & 13.77 +-  1.82\\
50926.8814 &  3.05 &16.75\nerr{2.49}{2.74} & -3.53\nerr{0.31}{0.38} &  7.66 +-  1.64\\
50927.8839 &  2.93 &30.35\nerr{3.16}{2.86} & -3.15\nerr{0.20}{0.22} & 13.34 +-  2.01\\
50928.8824 &  2.92 &16.75\nerr{2.49}{2.74} & -3.19\nerr{0.31}{0.35} &  9.69 +-  1.76\\
50929.8848 &  2.81 &34.18\nerr{3.56}{2.60} & -3.00\nerr{0.15}{0.17} & 18.54 +-  2.24\\
50930.8845 &  2.72 &37.00\nerr{3.05}{3.49} & -2.82\nerr{0.13}{0.15} & 23.62 +-  2.51\\
50947.9065 &  0.97 &18.14\nerr{4.41}{4.39} & -2.12\nerr{0.44}{0.55} & 13.11 +-  3.28\\
50948.8937 &  1.44 &13.48\nerr{3.28}{3.46} & -2.57\nerr{0.31}{0.41} & 11.43 +-  2.58\\
50956.8999 &  0.83 &29.75\nerr{6.52}{7.20} & -3.03\nerr{0.53}{0.63} & 16.20 +-  4.17\\
\hline
&&&&\\ 
\multicolumn{5}{l}{\footnotesize  $^a$ MJD}\\
\multicolumn{5}{l}{\footnotesize $^b$ duration of measurement in hours}\\
\multicolumn{5}{l}{\footnotesize $^c$ in ($10^{-12}\, \mbox{cm$^{-2}$ s$^{-1}$ TeV$^{-1}$}$)}\\
\multicolumn{5}{l}{\footnotesize $^d$ in ($10^{-12}\, \mbox{cm$^{-2}$ s$^{-1}$}$)}
\vspace*{-0.6cm}
\end{tabular}
\end{center}}
\end{table}
\subsection{The data of April 1998}
\begin{figure}
\resizebox{\hsize}{!}{\includegraphics{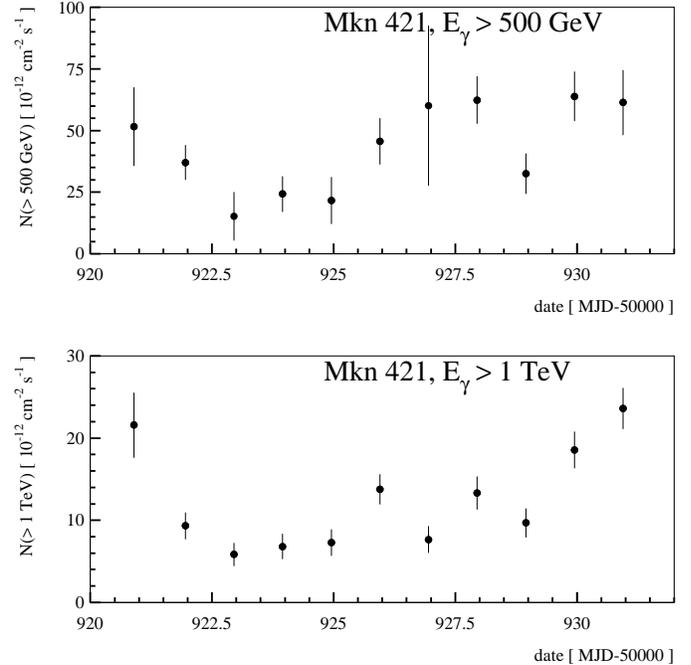}}
\vspace*{-0.5cm}
\caption{\label{mwc} \small
Integral HEGRA fluxes for Mkn 421 during the world-wide 
multiwavelength campaign 
for a threshold energy of 500~GeV in the upper panel and a 
threshold energy of 1~TeV in the lower panel. 
(The statistical errors of the 500~GeV data sample are larger than for the
1~TeV data sample, since events near the detector threshold of 500 GeV,
which were taken under larger zenith angles 
enter the integral flux with large weights.)}
\vspace*{-0.5cm}
\end{figure}
\begin{figure}
\vspace*{-0.5cm}
\resizebox{\hsize}{!}{\includegraphics{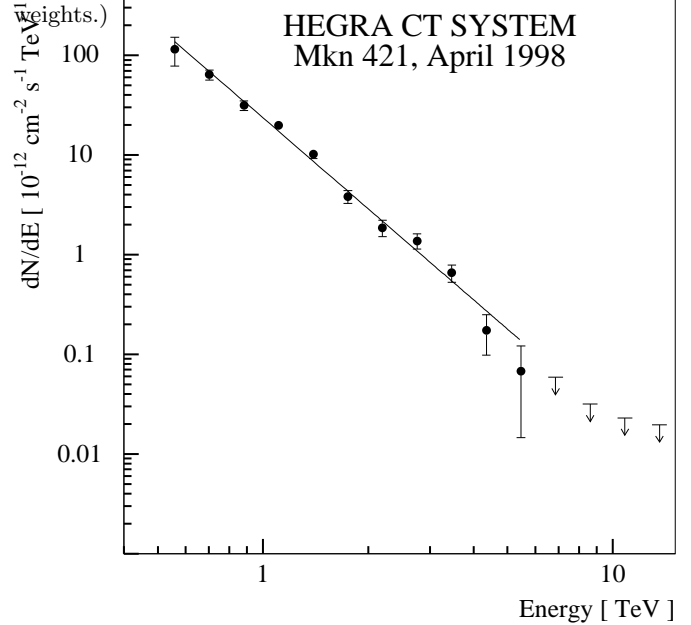}}
\vspace*{-0.5cm}
\caption{\label{aprilspec} \small The Mkn 421 spectrum during the April 1998
 observation period (17.-28.04.1998). The solid line shows a power law fit. 
Upper limits are at 2$\sigma$ confidence level.
(statistical errors only;
systematic errors as for Fig.\ \protect\ref{spec01})}
\vspace*{-0.5cm}
\end{figure}
As mentioned above, the data of April 1998 are 
of special interest since a world-wide
multiwavelength campaign was performed 
with measurements in the radio, optical, 
and UV regimes, at X-Ray energies (RXTE, BEPPO Sax, ASCA), 
at GeV energies, and at TeV energies.

In Figure \ref{mwc}, the HEGRA integral fluxes above 500 GeV and above 
1 TeV are shown (data up to 30$^\circ$ zenith angle in both plots). 
The flux varied by a factor of up to four.
Note, that all days of the multiwavelength campaign are included in Table 
\ref{strong}, since the emission was  high and the observation 
windows were large.

In Fig.\ \ref{aprilspec}, the mean spectrum during the April campaign is shown.
A power law fit yields a differential
spectral index of -3.03 $\pm0.08_{\rm stat}\, \pm0.10_{\rm syst}$.\newline
\clearpage
\begin{figure*}
\resizebox{\hsize}{!}{\includegraphics{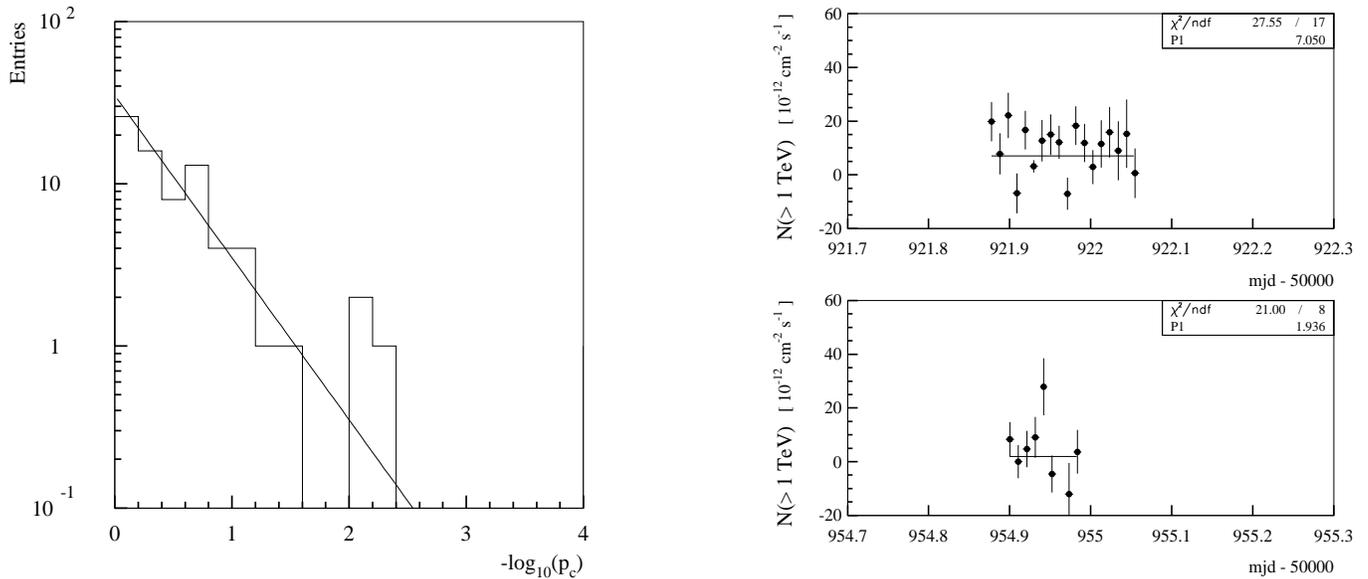}}
\caption{\label{tvarint2} \small The left side shows the distribution of the chance
probabilities
for finding for each night a stronger variability than observed, as computed for the
1 TeV threshold data (selection B). 
The right side shows for two nights the integral flux above 1 TeV
as a function of time. Each value has been determined with a 15 minute bin.}
\end{figure*}
The mean differential flux at 1 TeV was (23.5 $\pm$1.1$_{\rm stat}$ 
$\pm$8.4$_{\rm syst}$)$10^{-12}$ 
$\rm cm^{-2} s^{-1} TeV^{-1}$, which roughly equals the Crab flux at 1 TeV.
The $\chi^2$-value computed with the statistical errors
is  16.1 for 9 degrees of freedom, corresponding to a chance probability for 
larger $\chi^2$-values of 6.5\%.
\subsection{Search for variability within individual nights}
Temporal variability within individual nights has been searched for using
the integral fluxes above 500 GeV (data with zenith angles up to 30$^\circ$) 
and above 1 TeV (data with zenith angles up to 45$^\circ$).
For the data of each night, the integral fluxes were determined with 15 min bins.
For the first data selection the mean observation duration is 1.7 h
per night, for the second one
it is 25\% higher, since data of larger zenith angles are included. 
Variability was searched for with a $\chi^2$-analysis, 
fitting constant models to the
integral flux rates and determining the chance probabilities 
$P_{\rm c}$ for the resulting $\chi^2$-values. 

The results for the 1 TeV energy threshold data sample are 
given in Fig.\ \ref{tvarint2}.
On the left side the distribution of the 
-log$_{10}(P_{\rm c})$-values is shown.
The chance probabilities distribute as expected in the absence 
of variability.
On the side of small chance probabilities with -log$_{10}(P_{\rm c})>2$,
a marginally significant excess of three nights above an expectation of 
0.76 can be recognized. 
Taking into account that 76 nights have been studied, 
the chance probability for finding such three nights is computed to be 4\%.
The integral fluxes of two nights are shown in 
Fig.\ \ref{tvarint2} (right side).
The lower panel shows one of the three 
nights with the most significant variability 
(May 21st/22nd, 1998, MJD 50954/50955). 
The examples exemplify that due to limited photon statistics, 
the search is only sensitive to strong flares.
Also for the 500 GeV energy threshold data sample, no individual
night with significant variability has been found.

A preliminary analysis of Whipple data taken during the
world-wide multiwavelength campaign indicated flux variability at
energies above 2 TeV on a time scale of several hours (Maraschi et al.\ 
\cite{Mara:99}). No simultaneous data was
taken with the IACT system of HEGRA.
\section{Discussion}
\label{SU}
In this paper we present detailed temporal and spectral
information about Mkn 421 during 1997 and 1998.
The mean 1997--1998 differential flux at 1~TeV was about 
one half of the flux of the Crab nebula. The light curve shows several
distinct flares with mean durations of typically one or two days and 
fluxes at 1~TeV of approximately the Crab level.
During the world-wide multiwavelength campaign in 1998, we 
observed a 10 day phase of increased Mkn 421
activity with a differential flux at 1 TeV of one Crab level.

The time averaged energy spectrum follows to good 
approximation a power law with a differential 
spectral index (500 GeV to 7 TeV) of 
-3.09$\pm$0.07$_{\rm stat}$~$\pm$0.10$_{\rm syst}$.
Our result is consistent with earlier measurements of the 
differential spectral index during periods of moderate flux, 
i.e.\ -3.6$\pm$1.0$_{\rm stat}$ 
(at $\ge$1.2TeV) 
measured during the 1994/1995 observation period with the
HEGRA CT1 telescope (Petry et al.\ 
\cite{Petr:96}) and -2.92~$\pm$0.22$_{\rm stat}$~$\pm$0.1$_{\rm syst}$ 
measured during the 1995/1996 observation period with the Whipple telescope 
(Zweerink et al.\ \cite{Zwee:97}).

We did not find highly significant deviations from this mean spectrum, 
neither by dividing the data in a 1997 and a 1998 data sample, nor by dividing the
data in a low and a high emission data set. Moreover, 
all diurnal spectra -- some can be determined with
an accuracy of 0.2 in the differential spectral index -- 
are consistent with the mean spectrum.

The Whipple group reported on an extraordinary Mkn 421 flare observed 
on May 7th, 1996 with an integral TeV flux of more than 5 Crab units and a flux 
increase by a factor of $\sim$2 within approximetaly 1~h
(Gaidos et al.\ \cite{Gaid:96}), which
permitted to determine a differential spectral index 
during the flare with good statistical accuracy
(Zweerink et al.\ \cite{Zwee:97}; Krennrich et al.\ \cite{Kren:99}).
An index of $-2.56\,\pm$0.07$_{\rm stat}$~$\pm0.1_{\rm syst}$
indicated a harder spectrum in the flaring state than in the quiescent 
state.
Whereas our Mkn 421 spectra are consistent with a spectral index, which is independent
of time and absolute flux, just as in the case of Mkn 501 (Paper 1),
the Whipple spectrum during the extraordinary flare appears to be harder.
Nevertheless, taking into account the statistical and systematic errors on the 
estimates of the photon power law index, the effect of spectral 
hardening during the very strong flare is only marginally significant.

In our data we did not find strong evidence for flux-variability 
within individual nights, although the sensitivity for this search was limited, 
due to short observation windows of approximately 2 h and the low levels of emission.

%
%
Due to the similar redshifts of Mkn 421 and Mkn 501, the comparison of 
their spectra promises
insights into the intergalactic absorption by the DEBRA.
\begin{figure}
\resizebox{\hsize}{!}{\includegraphics{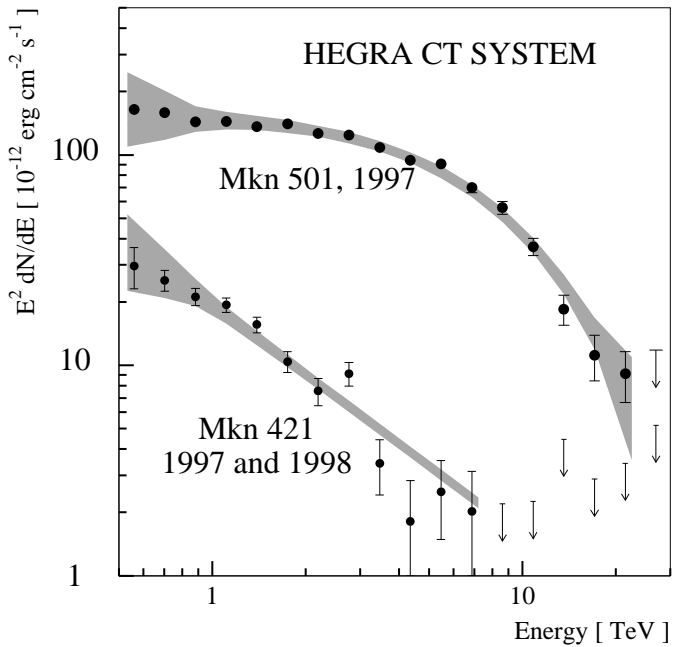}}
\vspace*{-0.5cm}
\caption{\label{comparison} \small The 1997 Mkn 501 and the 1997--1998 Mkn 421 
spectral energy distributions. The hatched regions show the systematic errors
on the shape of the spectra (see text).
Upper limits are at 2$\sigma$ confidence level.
}
\vspace*{0.5cm}
\end{figure}
In Fig.\ \ref{comparison}, the 1997--1998 Mkn 421 spectrum 
is compared to the 1997 Mkn 501 spectrum. 
During 1997 Mkn 501 was on average at 1 TeV more than 7 times 
brighter than Mkn 421 during 1997--1998.
The Mkn 421 spectrum is much softer than the Mkn 501
spectrum. Between energies of one and several TeV 
the differential spectral index is -3.1 for Mkn 421 and -2.2 for Mkn 501.
Certainly, due to the similar redshifts of both objects, the difference has to
be attributed to different intrinsic spectra and not to differences in the extinction 
by the DEBRA.
In the Mkn 501 spectrum we recorded an exponential cutoff at 
6.2~TeV, which could well be the result of the intergalactic absorption
by the DEBRA (see Paper 2 for a detailed discussion of possible origins 
of the exponential cutoff).
As can be recognized from the fits to the Mkn 421 data
of Eqs. \ref{pow} and \ref{exp}, we cannot exclude an exponential cutoff
for cutoff energies above $\simeq$2.8 TeV; this question will 
further be studied in future observations.
In this presentation the Mkn 421 spectrum
shows some curvature around 1 TeV. However, taking into account the
systematic errors on the shape of the spectrum (see hatched regions 
in Fig.\ \ref{comparison}) the curvature is surely not significant.

We anticipate that detailed modeling of the temporal and spectral 
characteristics of Mkn 421 during the multiwavelength campaign, 
combining the TeV data with
informations from longer wavelengths, will permit to gain 
important constraints on the astrophysical conditions inside the jet.\\[1ex]

{\it Acknowledgments}. 
The support of the German ministry for Research and
technology BMBF and of the Spanish Research Council CYCIT is gratefully
acknowledged. We thank the Instituto de Astrophysica de Canarias
for the use of the site and for supplying excellent working conditions at
La Palma. We gratefully acknowledge the technical support staff of the
Heidelberg, Kiel, Munich, and Yerevan Institutes. 
We thank T. Takahashi, E. Pian, M. Urry, and R. Remillard for
the coordination of the multiwavelength campaign, so that a substantial 
part of the data presented in this paper was taken simultaneously 
with X-Ray data.


\begin{thebibliography}{}
\bibitem[1999a]{Ahar:99a} {Aharonian F.A., 
Akhperjanian A.G., Barrio J.A., Bernl\"ohr K., Bojahr H., et al.,  
1999a, A\&A 342, 69 ({\it Paper 1})}
\bibitem[1999b]{Ahar:99b} {Aharonian F.A., 
Akhperjanian A.G., Barrio J.A., Bernl\"ohr K., Bojahr H., et al.,  
1999b, submitted to A\&A ({\it Paper 2})}
\bibitem[1999c]{Ahar:99c} {Aharonian F.A.,
Akhperjanian A.G., Barrio J.A., Bernl\"ohr K., Bojahr H., et al.,
1999b, submitted to A\&A, astro-ph/9901284}
\bibitem[1999]
{Bedn:99}{Bednarek W., Protheroe R.J., 1999, submitted to MNRAS}
\bibitem[1995]{Blan:95}{Blandford R.D, Levinson A., 1995, ApJ 441,79}
\bibitem[1993]{Bloo:93}{Bloom S.D., Marscher A.P., 1993. In: AIP
Conf.\ Proc.\ 280, Compton Gamma-Ray Observatory, ed.\ 
Friedlander M., Gehrels N., Macomb D.J. (New York AIP), 578}
\bibitem[1997]{Brad:97}
{Bradbury S.M., Deckers T., Petry D., 
Konopelko, A., Aharonian F., et al., 1997, A\&A 320, L5}
\bibitem[1998]{bul:98}{Bulian N., Daum A., Hermann G., Hess M.,
Hofmann W., et al., 1998, Astropart.\ Phys.\ 8, 223}
\bibitem[1997]{Copp:97}{Coppi P.S., 1997. In:
Cracow Workshop on Relativistic Jets in AGNs, eds.
     Ostrowski M., Sikora M., Madejski G., and Begelman M., 
Jagellonian University Press, p.\ 333}
\bibitem[1999]{Copp:99}{Coppi P.S. \& Aharonian F.A., 1998,
ApJ Let., submitted} 
\bibitem[1997]{Daum:97}{Daum A., Hermann G., He{\ss} M., 
Hofmann W., Lampeitl H., et al., 1997, Astropart.\ Phys.\ 8, 1}
\bibitem[1994]{Derm:94}{Dermer C.D \& Schlickeiser R., 1994, Ap.J Suppl. 90, 945}
\bibitem[1998]{Foss:98}{Fossati G., Maraschi L., Celotti A., 
Comastri A., Ghisellini G., 1998, MNRAS 299, 433}
\bibitem[1997]{Fras:97} {Fra{\ss} A., K\"ohler C., Hermann G., He{\ss} M., Hofmann W., 1997, Astropart.\ Phys.\ 8, 
91}
\bibitem[1996]
{GhisMD:1996}{Ghisellini G., Maraschi L., Dondi L., 1996, A\&AS 120, 503}
\bibitem[1965]
{Goul:65}{Gould J. \& Schreder, 1965, Phys. Rev. Lett. 16, 252}
\bibitem[1999]{Hart:99}{Hartman R.C.,Bertsch D.L., Bloom S.D., Chen A.W.,
Deines-Jones P., et al., 1999, accepted for publication in ApJS}
\bibitem[1998]{Hemb:98}{Hemberger M., 1998, Ph.D. thesis,
Heidelberg}
\bibitem[1996]{Gaid:96}{Gaidos J.A., Akerlof C.W., Biller S.D., 
Boyle P. J., Breslin A.C., et al., 1996, Nat 383, 319}
\bibitem[1997]{her:95}{Hermann G., 1995. In: Procs.
Towards a Major Atmospheric Cherenkov Detector~IV,
M. Cresti (ed), Padova, p. 396}
\bibitem[1997]{hof:97}{Hofmann W., 1997. In: Procs.
Towards a Major Atmospheric Cherenkov Detector~V,
De Jager O.C. (ed), Kruger Park, South Africa, p. 284}
\bibitem[1998]{kon:98}{Konopelko A. (HEGRA collaboration), 1998. In
Proc. 14th Europ. Cosmic Ray Symp., Madrid, in press}
\bibitem[1999]{Kono:99}{Konopelko A., Hemberger M., Aharonian F.,
Akhperjanian A.G., Barrio J.A., et al.\, 1999, 
accepted for publication in Astrop.\ Phys.}
\bibitem[1999]{Kren:99}{Krennrich F., Biller S.D., Bond I.H.,
Boyle P.J., Bradbury S.M., 1999, ApJ 511, 149}
\bibitem[1993]{Mann:93}{Mannheim K., 1993, A\&A 269, 67}
\bibitem[1999]{Mara:99}{Maraschi L., Fossati G., 
Tavecchio F., Chiappetti L., Celotti A.,et al., 1999.\
In: Proc.\ the Veritas Workshop on the TeV
Astrophysics of Extragalactic Objects, 
ed. Weekes T.C., Catanese M., Astroparticle Physics in press.}
\bibitem[1997]{Mast:97}{Mastichiadis A., Kirk J.G., 1997, A\&A 320, 19}
\bibitem[1994]{Mirz:94}{Mirzoyan R., Kankanian R., Krennrich F., et al., 1994, NIM A 315, 513}
\bibitem[1996]{Petr:96}{Petry D., Bradbury S.M., Konopelko A., 
Fernandez J., Aharonian F., et al., 1996, A\&A 311, L13}
\bibitem[1997]{Prot:98}{Protheroe R.J.,
Bhat C.L., Fleury P., Lorenz E., Teshima M., Weekes T.C., 1998. In 
Proc. 25th ICRC, Durban, vol. 8, p.317}
\bibitem[1997]{pul:97}{P\"uhlhofer G.,  Daum A.,
Hermann G., Hess M., Hofmann W.,
et al., 1997, Astropart. Phys.\ 8, 101}
\bibitem[1992]{Punc:92}{Punch M., Akerlof C.W., Cawley M.F., 
Chantell M., Fegan D.J., et al., 1992, Nat 358, 477}
\bibitem[1996]{Quin:96}
{Quinn J., Akerlof C.W., Biller S.,
Buckley J., Carter-Lewis D.A., et al., 1996, ApJ 456, L83}
\bibitem[1995]{Raut:95} Rauterberg G., M\"uller N., Deckers T. (HEGRA collaboration), 
1995. In: Proc. 24th ICRC, Rome, 3, 460
\bibitem[1994]{Siko:94}{Sikora M., Begelman M.C., Rees M.J.,
1994, ApJ 421, 153}
\bibitem[1997]{Siko:97}{Sikora M., 1997,
In: Proc. 4th Compton Symposium, AIP Conf. Proc.,
Dermer C., Strickman M., Kurfess J. (eds), 494}
\bibitem[1992]
{Stec:92}{Stecker F.W., De Jager O.C. \& Salamon M.H.,1992, ApJ.\ 390, L49}
\bibitem[1999]{Taka:99}{Takahashi T., Madejski G., Kubo H., 1999.\
In: Proc.\ the Veritas Workshop on the TeV
Astrophysics of Extragalactic Objects, 
ed. Weekes T.C., Catanese M., Astroparticle Physics in press.}
\bibitem[1997]{umu:97}{Ulrich M.H., Maraschi L.,
Urry C.M., 1997, ARA\&A 35, 445}
\bibitem[1999]{Urry:99}{Urry C.M., 1999.\
In: Proc.\ the Veritas Workshop on the TeV
Astrophysics of Extragalactic Objects, 
ed. Weekes T.C., Catanese M., Astrop.\ Physics in press.}
\bibitem[1997]{Zwee:97}{Zweerink J.A., Akerlof C.W.,
Biller S.D., Boyle P., Buckley J.H., et al., 1997, ApJ 490, L141}
\end{thebibliography}
\end{document}